\documentclass[letter,aps,superscriptaddress,nofootinbib,tightenlines,prd]{revtex4}

\usepackage{amsfonts}
\usepackage{multirow}
\usepackage{mathrsfs}
\usepackage{graphicx}
\usepackage{amsmath}
\usepackage{amssymb}
\usepackage{bm}
\usepackage{bbm}
\usepackage{color}
\usepackage{ulem}
\usepackage{subfig}
\usepackage{caption}
\usepackage{graphicx}
\usepackage{epsfig}
\usepackage{braket}
\usepackage{slashed}
\usepackage{hyperref}
\usepackage{ulem}
\usepackage{makecell}

\newcommand{\msn}{{\pi^-}}
\newcommand{\bsq}{\begin{subequations}}
\newcommand{\esq}{\end{subequations}}

\newcommand{\nn}{\nonumber}

\newcommand{\beq}{\begin{equation}}
\newcommand{\eeq}{\end{equation}}
\newcommand{\bqa}{\begin{eqnarray}}
\newcommand{\eqa}{\end{eqnarray}}
\newcommand{\bseq}{\begin{subequations}}
\newcommand{\eseq}{\end{subequations}}

\newcommand{\QCDtwo}{${\rm QCD}_2$}

\newcommand{\non}{ \nonumber \\}

\def\Xint#1{\mathchoice
{\XXint\displaystyle\textstyle{#1}}%
{\XXint\textstyle\scriptstyle{#1}}%
{\XXint\scriptstyle\scriptscriptstyle{#1}}%
{\XXint\scriptscriptstyle\scriptscriptstyle{#1}}%
\!\int}
\def\XXint#1#2#3{{\setbox0=\hbox{$#1{#2#3}{\int}$ }
\vcenter{\hbox{$#2#3$ }}\kern-.6\wd0}}

\def\dashint{\Xint-}
\def\massu{{\sqrt{2\lambda}}}

\begin{document}

\title{\mbox{}\\[14pt]
Strange-antistrange and charm-anticharm asymmetries of pion in 't Hooft model}

\author{Mingliang Zhu~\footnote{lightzhu@csu.edu.cn}}
\affiliation{School of Physics, Central South University, Changsha 410012, China\vspace{0.2 cm}}

\author{Siwei Hu~\footnote{husw@ihep.ac.cn}}
\affiliation{Institute of High Energy Physics, Chinese Academy of Sciences, Beijing 100049, China\vspace{0.2 cm}}
\affiliation{School of Physical Sciences, University of Chinese Academy of Sciences,
	Beijing 100049, China\vspace{0.2 cm}}

\author{Yu Jia~\footnote{jiay@ihep.ac.cn}}
\affiliation{Institute of High Energy Physics, Chinese Academy of
Sciences, Beijing 100049, China\vspace{0.2 cm}}
\affiliation{School of  Physical Sciences, University of Chinese Academy of Sciences,
Beijing 100049, China\vspace{0.2 cm}}

\author{Zhewen Mo~\footnote{mozw@itp.ac.cn}}
\affiliation{Institute of Theoretical Physics, Chinese Academy of Sciences, Beijing 100190, China\vspace{0.2 cm}}
\affiliation{Institute of High Energy Physics, Chinese Academy of
Sciences, Beijing 100049, China\vspace{0.2 cm}}

\author{Xiaonu Xiong~\footnote{xnxiong@csu.edu.cn}}
\affiliation{School of Physics, Central South University, Changsha 410012, China\vspace{0.2 cm}}

\begin{abstract}
As a sequel of our preceding work [S.~Hu {\it{et al.}}, Phys.~Rev.~D 108 (2023) 9, 094040], we investigate the strange-antistrange
and charm-anticharm asymmetries in the parton distribution functions (PDFs) of a light flavored meson, exemplified by
the first excited pion in the 't Hooft model, {\it viz.}, QCD in two spacetime dimensions with infinite number of colors.
Counted as an ${\cal O}(1/N_c)$ effect, the intrinsic strange content necessarily originates from the higher Fock component of the light flavored meson, 
which entails infinite towers of $K$ and $\overline{K}$ mesons. 
Numerical studies reveal that, with $m_u/m_d=1/2$, the $s$-$\bar{s}$ and $c$-$\bar{c}$ asymmetries of the first excited $\pi^-$
can reach per cents level. While the $s$-$\bar{s}$ asymmetry predicted from the meson cloud model (MCM) grossly align with the rigorous approach, 
there exists severe discrepancy between two approaches on the $c$-$\bar{c}$ asymmetry.
\end{abstract}

\date{\today}

\maketitle

\section{Introduction}

The parton distribution functions (PDFs) encode the key characteristic of the internal nucleon structure,
which provide crucial nonperturbative inputs for all the QCD factorization theorem based predictions
in the high energy $pp$, $ep$ collision experiments.  Our current knowledge on the proton PDFs has entered the precision era,
largely facilitated by the data-driven global fitting recipe~\cite{Martin:2009iq, Hou:2019efy, Ablat:2024muy, NNPDF:2021njg, NNPDF:2024dpb}, and by the lattice calculation~\cite{Hobbs:2019gob,Constantinou:2020hdm,Ji:2020ect}. %and recently by the direct calculation on Euclidean lattice and
To date we have already gleaned a wealth of robust knowledge about the valence quark distributions inside nucleon.
In contrast, our understanding about the sea quark content of a nucleon, say, the $s$ and $c$ quarks,
still remains elusive.

Stimulated by the anomalous value of the Weinberg angle extracted from the $\nu N$ collision experiment
by {\tt NuTeV} Collaboration~\cite{NuTeV:2001whx,NuTeV:2001dfo,NuTeV:2002ryj}, a flurry of theoretical work has emerged that speculate on the possible $s$-$\bar{s}$ asymmetries in proton PDF~\cite{Brodsky:1996hc,Wang:2016ndh,Du:2017nzy,Salajegheh:2015xoa,Vega:2015hti}.
The sea quarks like $s$, $\bar{s}$ can either arise from the gluon splitting, or from the higher Fock component of the proton,
{\it viz.}, $\vert u u d s \bar{s} \rangle$.
Through some careful pQCD analysis, the gluon splitting mechanism was estimated
to merely generate very tiny $s$-$\bar{s}$ asymmetry~\cite{Catani:2004nc}.
Hence the potential $s$-$\bar{s}$ asymmetry is widely believed to be largely attributed to the higher Fock state of the proton.
A popular phenomenological model dubbed the meson cloud model (MCM)\cite{Sullivan:1971kd}, in which the the dominant higher Fock component of the proton
may be approximated as the $\Lambda$ baryon surrounded by a kaon cloud, allows one to make some quantitative predictions to the $s$-$\bar{s}$ asymmetry
inside the proton.

The intrinsic heavy quark content of nucleon, such as the intrinsic charm,  has also been conjectured to be non-negligible
by Brodsky and Hoyer long ago~\cite{Brodsky:1980pb,Brodsky:1981se}.
Nevertheless, in contrast to intrinsic strange, it is widely believed that the intrinsic charm content is highly suppressed,
roughly with a speed $\propto 1/m_c^2$~\cite{Franz:2000ee}.
Recently there have been revived interests toward the intrinsic charm PDF, largely propelled by the
$Z+c$ jet measurement by {\tt LHC}~\cite{LHCb:2021stx} and the latest nucleon PDF released
by {\tt NNPDF} Collaboration~\cite{Ball:2022qks, NNPDF:2023tyk}.

To deepen our knowledge about the nucleon structure, it is compulsive to build some robust understanding toward the intrinsic strange and charm PDFs.
While lattice QCD will definitely play a vital role to eventually pin down the intrinsic sea quark distributions and the pertaining asymmetries,
it is also desirable to look at this problem from some alternative perspective, in which some transparent physical picture can be drawn
and some analytical understanding may be achieved.

It turns out to be useful to learn some lessons about hadron structure from the solvable field theory models.
The 't Hooft model~\cite{tHooft:1974pnl}, the two-dimensional QCD in the $N_c\to\infty$ limit, has proven to be a fruitful laboratory to meet such a purpose,
since it shares some common features as the realistic QCD in four dimensions, such as color confinement and ``spontaneous" chiral symmetry breaking.
During the past decades, a plethora of works have been devoted to explore various hadronic phenomena in 't Hooft model~\cite{Callan:1975ps,Einhorn:1976uz,Einhorn:1976ax,Brower:1977hx,Brower:1977as,Brower:1978wm}.

Employing the light-front quantization and working at the ${\cal O}(1/N_c)$ accuracy, recently
we have investigated the intrinsic charm PDF in a light flavor neutral meson in the 't Hooft model~\cite{Hu:2022wsf}.
It is found that the MCM predictions for intrinsic charm PDF in ${\rm QCD}_2$ significantly differs from the rigourous field-theoretical predictions.
Since the ${\rm QCD}_2$ considered in \cite{Hu:2022wsf} only entails a single flavor of light quark,
thus there is no $c$-$\bar{c}$ asymmetry in the light flavor neutral meson owing to charge conjugation symmetry.

In this work, we extend the preceding work~\cite{Hu:2022wsf} to consider a situation closer to the realistic ${\rm QCD}_4$,
in which two lightest quarks $u$, and $d$ satisfy $m_u/m_d\approx 1/2$, and the lightest meson is the pseudo-Goldstone particle, ``pion".
The strange and charm quarks are assumed to be much heavier than the $u$ and $d$.
It is conceivable that the $s$-$\bar{s}$ and $c$-$\bar{c}$ asymmetries would be generated
by the isospin breaking effect once including the higher Fock component of pion.
We compare the numerical predictions from the rigourous field-theoretical calculation and MCM
for both $s$-$\bar{s}$ and $c$-$\bar{c}$ asymmetries inside the $\pi^-$.

As a caveat, it should be warned that the underlying mechanism for color confinement and spontaneous symmetry breaking
is utterly different between 't Hooft model and realistic ${\rm QCD}_4$. In particular, the gluon fields in the former are secretly non-dynamical degrees of freedom,
due to the absence of transverse space.
However, it is still rewarding to use this solvable model to critically test the reliability of some influential phenomenological models,
otherwise hardly feasible in the realistic ${\rm QCD}_4$.
Concretely speaking, the central goal of this work is to examine to which extent, the MCM predictions for $s$-$\bar{s}$ and $c$-$\bar{c}$ asymmetries
in a pion align with the first principle predictions.

The rest of the paper is distributed as follows.
%-------------------
In Sec.~\ref{setup:stage} we set up the stage by briefly reviewing the Hamiltonian approach
in light-front quantization in \QCDtwo.
%-------------------
In Sec.~\ref{sec:icpdf} we present a rigorous derivation of the strange and antistrange PDFs of a light flavored meson in 't Hooft model,
accurate at ${\cal O}(1/N_c)$. The respective PDFs are expressed
in terms of the convolution of the mesonic light-cone wave functions.
%-------------------
In Sec.~\ref{sec:MCM}, we present a derivation of the $s$ and $\bar{s}$ PDFs of a light flavored meson in the spirit of the meson cloud model.
%-------------------
In Sec.~\ref{sec:numerical}, we first explain the recipe of setting the light quark masses, then
present the numerical results for the intrinsic $s$ and $\bar{s}$ PDFs of the first excited $\pi^-$ and the respective asymmetries.
To gauge how the asymmetry depends on the sea quark mass, we also investigate the $c$-$\bar{c}$ asymmetry of the first excited $\pi^-$.
%-------------------
 Finally we summarize in Sec.~\ref{sec:summary}.
%-------------------

\section{Set up the stage}
\label{setup:stage}

Let us concentrate on the situation $m_u\sim m_d\ll m_s$, where
the corresponding ${\rm QCD}_2$ Lagrangian reads
%-------------------
\beq
%-------------------
\mathcal{L}_{{\rm QCD}_2}= \sum_{q=u,d,s} \overline{q}\left(i\slashed{D}-m_q\right)q - \frac{1}{4}F^{a,\mu\nu}F^{a}_{\mu\nu},
%-------------------
 \label{QCD:lagr}
%-------------------
\eeq
%-------------------
where $D_\mu= \partial_\mu-ig_s A_\mu^aT^a$ signifies the color covariant derivative
and $T^a$ denotes the generators of the $SU(N_c)$ group in the fundamental representation.
The gluon field strength tensor is defined as $F_{\mu\nu}^a \equiv \partial_\mu A_\nu^a-\partial_\nu A_\mu^a+g_sf^{abc}A_\mu^bA_\nu^c$.
For convenience, the chiral-Weyl representation for the Dirac $\gamma$ matrices is used:
%---------------------------
\begin{equation}
    \gamma^0=\sigma_1,\quad \gamma^1=-i\sigma_2,\quad \gamma_5\equiv \gamma^0\gamma^1=\sigma_3,
\end{equation}
%---------------------------
where $\sigma_{i}(i=1,2,3)$ are the Pauli matrices.

The quark spinor field can be decomposed into
$   q = 2^{-{\frac{1}{4}}}
    \left(
    \begin{array}{c}
    q_R \\ q_L
    \end{array}
    \right)$,
with $q_{R/L} = \frac{1 \pm \gamma_5}{2} q$ indicatng the right-handed and left-handed components.

In this work we are also interested in the $N_c \to \infty$ limit, specified by the condition
%-------------------
\beq
%-------------------
    N_c \to \infty, \qquad \qquad \lambda\equiv\frac{g^{2}N_c}{4\pi}\;\, \text{fixed}.
%-------------------
\eeq
%-------------------
The 't Hooft coupling $\lambda$ sets the characteristic hadronic scale in ${\rm QCD}_{2}$, analogous to
$\Lambda_{\rm QCD}$ in the realistic ${\rm QCD}_{4}$.

Adopting the light-cone coordinates  $x^\pm= x_{\mp}= (x^0\pm x^1)/\sqrt{2}$ and impose the light-cone gauge $A^{+\,a} = 0$,
one finds that $q_L$ and $A^{-\,a}$ are non-propagating auxiliary fields upon using Euler-Lagrangian equation, and the light-front
Hamiltonian can be built out of the canonical variables $q_R^\dagger$ and $q_R$. After equal light-front time quantization,
the canonical variable $q_R$ can be expanded in terms of quark annihilation and creation operators:
%-------------------
\beq
%-------------------
q_R^{i} = \int_0^\infty \frac{dk^+}{2\pi} \left[ b^i(k^+) e^{-ik^+x^-} + d^{i\dagger}(k^+)e^{ik^+x^-}\right],
%-------------------
\label{right:handed:quark:field:Fourier:transf}
%-------------------
\eeq
%-------------------
with $i=1,\cdots, N_c$ indicating the color index.

The technique of bosonization~\cite{Kikkawa:1980dc,Nakamura:1981zi,Rajeev:1994tr,Dhar:1994ib,Dhar:1994aw,Cavicchi:1993jh,Barbon:1994au,Itakura:1996bk}
turns out to be useful to diagonalize the light-front Hamiltonian.
One can define a set of color-singlet compound operators $M$, $B$ and $D$ from the quark/antiquark creation and annihilation operators.
For example, one defines $M$ as
%-------------------
 \beq
%-------------------
M^{\bar f_1f_2}\left(k^{+}, p^{+}\right) =
\frac{1}{\sqrt{N_c}} \sum_{i} d_i^{f_1}(k^{+})b_i^{f_2}(p^{+}),
%-------------------
\eeq
%-------------------
which is subject to the commutation relation:
%-------------------
\beq
%-------------------
    [M^{\bar f_1f_2}(k_1^{+},p_1^{+}), M^{\dagger\bar f_3f_4}(k_2^{+},p_2^{+})]
    =(2\pi)^{2}
    \delta_{f_1f_3}\delta_{f_2f_4}
    \delta(k_1^{+}-k_2^{+})\delta(p_1^{+}-p_2^{+})+\mathcal{O}\left(\frac{1}{N_c}\right).
%-------------------
\eeq
%-------------------

In a color confining theory,  it is the color-singlet quark-antiquark pair, rather than the isolated quark or antiquark,
that can be created or annihilated in a physical process. As a consequence,
the compound operators $B$ and $D$ are not independent, instead can be expressed as the
convolution of the $M^\dagger M$.  Henceforth one reaches a LF Hamiltonian build out of the integral over $M^\dagger M$.

In order to diagonalize the LF Hamiltonian, one further trade the compound operators $M$ and $M^\dagger$
for the mesonic annihilation and creation operators $m_n$ and $m_n^\dagger$ ($n$ signifies the $n$-th excited meson)
through the relation:
%---------------------------
\beq
%---------------------------
M^{\bar{f}_1f_2}((1-x)P^{+},xP^{+}) = \sqrt{\frac{2\pi}{P^{+}}}
\sum_{n=0}^{\infty}\varphi^{f_2\bar f_1}_{n}(x)m^{f_2\bar f_1}_{n}(P^{+}),
%---------------------------
\label{M to m}
%---------------------------
\eeq
%---------------------------
where the coefficient function $\varphi^{f_1\bar{f}_2}_n(x)$ is interpreted as the light-cone wave function (LCWF) of the $n$-th excited meson with the flavor
content $f_1\overline{f}_2$.

It is natural to demand that the mesonic annihilation and creation operators obey the standard commutation relation:
%---------------------------
\beq
%---------------------------
    \left[m^{f_i\bar f_j}_n(P^+_1), {m^{\dagger f_k\bar f_l}_r}(P^+_2)\right]
    =
    2\pi\delta_{f_if_k}\delta_{f_jf_l}\delta_{nr}\delta(P^+_1-P^+_2)
    + \mathcal{O}\left(\frac{1}{N_c}\right).
%---------------------------
\label{m_commute}
%---------------------------
\eeq
%---------------------------

To comply with \eqref{m_commute}, the LCWFs must satisfy the following orthogonality and completeness conditions:
%---------------------------
\begin{subequations}
%---------------------------
\bqa
%---------------------------
&& \int^{1}_{0} dx\,
    \varphi^{f_1\bar f_2}_{n}(x)\varphi^{f_1\bar f_2}_{m}(x)
=
\delta_{nm},
%---------------------------
\\
%---------------------------
&&    \sum_{n}
    \varphi^{f_1\bar f_2}_{n}(x)\varphi^{f_1\bar f_2}_{n}(y)
= \delta(x-y).
%---------------------------
\eqa
%---------------------------
\end{subequations}
%---------------------------

At the lowest order in $1/N_c$, the light-front Hamiltonian is expected to consist of
all possible free mesons:
%---------------------------
\beq
%---------------------------
    H_{\mathrm{LF}}= P^- =
    \sum_{n,f_1f_2}\int
    \frac{dP^{+}}{2\pi}P^{-}_{n,f_1f_2}
    m^{\dagger f_1\bar f_2}_{n}(P^{+})m^{f_1\bar f_2}_{n}(P^{+})
    + \mathcal{O}\left(\frac{1}{\sqrt{N_c}}\right),
%---------------------------
\label{eq:H_LF}
%---------------------------
\eeq
%---------------------------
where for simplicity we have suppressed the irrelevant vacuum energy piece.

In order to reach such a diagonalized form, the mesonic LCWF must obey the celebrated 't Hooft equation~\cite{tHooft:1974pnl}:
%---------------------------
\beq
%---------------------------
    \left(\frac{m_1^2-2\lambda}{x}+\frac{m_2^2-2\lambda}{1-x}\right)
      \varphi_n^{f_1\bar f_2}\left(x\right) -2\lambda
    \dashint_0^1 dy {\varphi_n^{f_1\bar f_2}(y) \over \left(x-y\right)^2} = \mu_{n,f_1,f_2}^2\varphi_n^{f_1\bar f_2}\left(x\right),
%---------------------------
\label{tHooft:equation}
%---------------------------
\eeq
%---------------------------
where $m_1$, $m_2$ are the current quark masses affiliated with flavor $f_1$ and $f_2$, respectively, $\mu_{n,f_1f_2}$ is the meson mass of the $n$-th excited mesonic state in the $f_1\bar{f}_2$ family.
The symbol $\dashint$ denotes the principal value (PV) prescription for an integral,
%---------------------------
\begin{equation}
%---------------------------
    \dashint dy\, \frac{f(y)}{(x-y)^2}
    =
    \lim_{\epsilon \rightarrow 0^+}
    \int dy \, \Theta(|x-y|-\epsilon)
    {\frac{f(y)}{(x-y)^2}  } - {\frac{2 f(x)}{\epsilon}  }.
%---------------------------
\end{equation}
%---------------------------

Since the intrinsic strange represents an ${\cal O}(1/N_c)$ correction, in the following we must extend \eqref{eq:H_LF} to
include the ${\cal O}(1/\sqrt{N_c})$ piece in the light-front Hamiltonian, which entails three meson coupling.
Consequently, we also need to numerically solve the 't Hooft equation for three distinct species of mesonic states:
$\pi^-(d\bar{u})$, $K^0(d\bar{s})$, and $K^-(s\bar{u})$.

\section{Rigorous expressions for $s$ and $\bar{s}$ PDFs of $\pi^-$ in 't Hooft model}
\label{sec:icpdf}

We start from the standard gauge-invariant operator definition for the $s$ quark PDF of a $\pi^-$~\cite{Collins:1981uw}:
%-------------------
\beq
%-------------------
    f_{s / \pi_n^-}(x) = \int \frac{dz^{-}}{4\pi}e^{-ixP^{+}z^{-}}\bra{\pi_n^-(P^+)}\overline
    s (z^{-})\gamma^{+}%\mathcal{W}[z^{-},0]
    \mathcal{P} \left[\exp\left(-ig_s\int^{z^-}_0 d\eta^- A^{+,a}(\eta^-)\right) T^a\right]
    s(0)\ket{\pi_n^-(P^+)}_{C},
%-------------------
\label{Def:strange:in:pion:PDF}
%-------------------
\eeq
%-------------------
where only the connected part is retained. The light-like Wilson line,
whose role is to ensure gauge invariance, can be simply dropped since
we are working with the light-cone gauge $A^{+,a}=0$.

Analogous to what is done in Ref.~\cite{Hu:2022wsf},
one can simplify the light-cone separated $s$ quark bilinear \eqref{Def:strange:in:pion:PDF} as $\overline{s}(z^{-})\gamma^+ s (0)=s^\dagger_{R}(z^{-})s_{R}(0)$.
Expanding the $s_R$ field in terms of $s$ annihilation and $\bar{s}$ creation operator as in \eqref{right:handed:quark:field:Fourier:transf},
and applying the bosonization procedure outlined in Sec.~\ref{setup:stage}, we can express the nonlocal strange quark bilinear
in terms of mesonic annihilation and creation operators $m_n$, $m_n^\dagger$:
%-------------------
\begin{align}
%-------------------
s^\dagger_{R}(z^{-})s_{R}(0) 	& =
	\int \frac{dk^+_1dk^+_2}{2\pi} {N_{c}}\delta(k^+_1-k^+_2)e^{-ik^+_1z^{-}}
	\non
%-------------------
	&  +\sum_{n}\int \frac{dk^+_1dk^+_2}{(4\pi)^{3/2}}\frac{\sqrt{N_{c}}}{\sqrt{k^+_1+k^+_2}}
	{e^{ik^+_1z^-}m^{\dagger s\bar{s}}_{n}}(k^+_1+k^+_2)
	{\varphi^{s\bar s}_{n}\left(\frac{k^+_1}{k^+_1\!+\!k^+_2}\right)}
%-------------------
\non
%-------------------
	&
	+\sum_{n}\int \frac{dk^+_1dk^+_2}{(4\pi)^{3/2}}\frac{\sqrt{N_{c}}}{\sqrt{k^+_1+k^+_2}}
	{e^{-ik^+_1z^-}m^{s\bar{s}}_{n}}(k^+_1+k^+_2)
	{\varphi^{s\bar s}_{n}\left(\frac{k^+_2}{k^+_1\!+\!k^+_2}\right)}
	\non
	&  +\sum_{f,n_1,n_2}\int \frac{dk^+_1dk^+_2dq^+}{(2\pi)^{2}}e^{ik^+_1z^{-}}
	{m^{\dagger s\bar{f}}_{n_1}}(k^+_1+q^+)m^{s\bar{f}}_{n_2}(k^+_2+q^+)
	\frac{\varphi^{s\bar{f}}_{n_1}\left( \frac{k^+_1}{k^+_1+q^+} \right)}{\sqrt{k^+_1+q^+}}
	\frac{\varphi^{s \bar{f}}_{n_2}\left( \frac{k^+_2}{k^+_2+q^+} \right)}{\sqrt{k^+_2+q^+}}\non
	& -\sum_{f,n_1,n_2}\int \frac{dk^+_1dk^+_2dq^+}{(2\pi)^{2}}e^{-ik^+_1z^{-}}
	{m^{\dagger f\bar{s}}_{n_1}}(k^+_2+q^+)m^{f\bar{s}}_{n_2}(k^+_1+q^+)
	\frac{\varphi^{f\bar{s}}_{n_1}\left( \frac{q^+}{k^+_2+q^+} \right)}{\sqrt{k^+_2+q^+}}
	\frac{\varphi^{f\bar{s}}_{n_2}\left( \frac{q^+}{k^+_1+q^+} \right)}{\sqrt{k^+_1+q^+}}.
%-------------------
\label{eq:bilinear}
%-------------------
\end{align}
%-------------------
As will be seen shortly, the first three lines, {\it viz.}, the ${\cal O}(N_c)$ and ${\cal O}(\sqrt{N_c})$ pieces,
do not contribute to the intended intrinsic strange PDF.

Next we turn to the higher Fock component of the physical $\pi^-$.
As mentioned before, in order to detect its intrinsic strange content, one has to extend the ${\rm QCD}_2$ light-front Hamiltonian
to next-leading order in $1/N_c$. The full light-front Hamiltonian can be split into $H_{{\text{LF}}}= H_{{\text{LF}},0}+V$, where the free mesonic Hamiltonian
$H_{{\text{LF}},0}$ is given in \eqref{eq:H_LF},
and the $V$ term encompasses all possible ${\cal O}(1/\sqrt{N_c})$ three-meson interactions.
Incorporating the first-order quantum-mechanical correction, the physical $\pi^-$ state can be expanded into
%---------------------------
\beq
%---------------------------
\ket{\msn'}
\approx \ket{\msn} + \frac{1}{P^--H_{\mathrm{LF},0}+i\epsilon}V \ket{\msn}.
%---------------------------
\label{NLO:Fock:component:pi:minus}
%---------------------------
\eeq
%---------------------------
$\ket{\msn'}$ denotes the eigenstate of the full LF Hamiltonian, and $\ket{\msn}$ signifies the eigenstate of $H_{{\text{LF}},0}$,
which can be generated by
%---------------------------
\beq
%---------------------------
    \ket{\pi^-_n(P^+)} = \sqrt{2P^+}m_n^{\dagger d\bar u}(P^+)\ket{0},
%---------------------------
\label{eq:pi_n:m}
%---------------------------
\eeq
%---------------------------
here $n$ denotes the principle quantum number.

The interaction potential $V$ accounts for the
the three-meson coupling, which scales as ${\cal O}(1/\sqrt{N_c})$~\cite{Callan:1975ps}.
To our concern, the most relevant parts in $V$  are those coupling $\pi^-$ with all possible excited $s\bar{u}$ and $d\bar{s}$ mesons:
%---------------------------
\beq
%---------------------------
V_{\rm strange}=\mathcal{V}+\overline{\mathcal{V}}+\mathrm{h.c.},
%---------------------------
\label{Vs}
%---------------------------
\eeq
%---------------------------
where
%---------------------------
\begin{subequations}
%---------------------------
\label{eq:mathcal_V}
%---------------------------
\begin{align}
%---------------------------
        \mathcal{V} =
        \frac{-\lambda}{(2\pi)^{ \frac{3}{2} }\sqrt{N_{c}}}
        &
        \sum_{n_1n_2n_3}
        \int_0^\infty dq^+ dk^+_1dk^+_2dk^+_3dk^+_4
        \delta(k^+_1\!-\!k^+_2\!+\!k^+_3\!+\!k^+_4)
        {m^{\dagger s\bar{u}}_{n_1}}(k^+_1\!+\!q^+)
        m^{d\bar{u}}_{n_2}(k^+_2\!+\!q^+)
        {m^{\dagger d\bar{s}}_{n_3}}(k^+_3\!+\!k^+_4) \nonumber\\
           &\ \ \ \ \ \ \ \
           \times\frac{1}{(k^+_3-k^+_2)^{2}}
           \frac{\varphi^{s\bar{u}}_{n_1}\left(\frac{k^+_1}{k^+_1\!+\!q^+}\right)}{\sqrt{k^+_1\!+\!q^+}}
           \frac{\varphi^{d\bar{u}}_{n_2}\left(\frac{k^+_2}{k^+_2\!+\!q^+}\right)}{\sqrt{k^+_2\!+\!q^+}}
           \frac{\varphi^{d \bar{s}}_{n_3}\left(\frac{k^+_3}{k^+_3\!+\!k^+_4}\right)}{\sqrt{k^+_3\!+\!k^+_4}},
 %---------------------------
 \\%  
 %---------------------------
    \overline{\mathcal{V}} =
        \frac{\lambda}{(2\pi)^{ \frac{3}{2} }\sqrt{N_{c}}}
        &
        \sum_{n_1n_2n_3}
        \int_0^\infty dq^+ dk^+_1dk^+_2dk^+_3dk^+_4
        \delta(k^+_1\!-\!k^+_2\!+\!k^+_3\!+\!k^+_4)
        {m^{\dagger d\bar{s}}_{n_1}}(k^+_1\!+\!q^+)
        m^{d\bar{u}}_{n_2}(k^+_2\!+\!q^+)
        {m^{\dagger s\bar{u}}_{n_3}}(k^+_3\!+\!k^+_4) \nonumber\\
           &\ \ \ \ \ \ \ \
           \times\frac{1}{(k^+_3-k^+_2)^{2}}
           \frac{\varphi^{d\bar{s}}_{n_1}\left(\frac{k^+_1}{k^+_1\!+\!q^+}\right)}{\sqrt{k^+_1\!+\!q^+}}
           \frac{\varphi^{d\bar{u}}_{n_2}\left(\frac{k^+_2}{k^+_2\!+\!q^+}\right)}{\sqrt{k^+_2\!+\!q^+}}
           \frac{\varphi^{s \bar{u}}_{n_3}\left(\frac{k^+_3}{k^+_3\!+\!k^+_4}\right)}{\sqrt{k^+_3\!+\!k^+_4}}.
%---------------------------
\end{align}
%---------------------------
\end{subequations}
%---------------------------

\begin{figure}[htbp]
        \centering
        \includegraphics[scale=1]{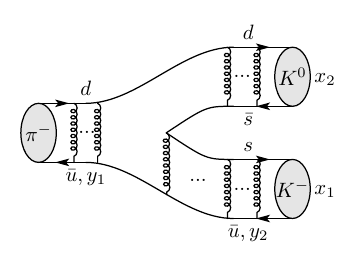}
        \includegraphics[scale=1]{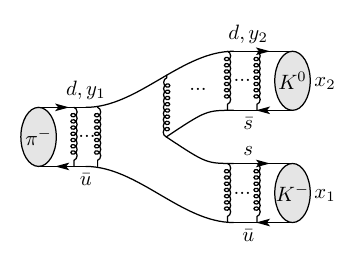}
    \caption{Pictorial illustration of the $\pi^-_n \to K^-_{n_1} + K^0_{n_2}$ transition  mediated by the triple meson vertex.
The dots represent arbitrary numbers of planar gluon exchanges.}
\label{fig:three_meson_vertex}
\end{figure}

To identify the next-to-leading order Fock component of $\pi_n^-$,
we insert a complete set of intermediate states in \eqref{NLO:Fock:component:pi:minus},
which are necessarily composed of the infinite towers of $K^0$ and  $K^-$.
The encountered transition matrix element for $\pi^-_n \to K^-_{n_1} + K^0_{n_2}$ can be expressed as
%---------------------------
\begin{align}
%---------------------------
\label{LS:matrix:element}
%---------------------------
\bra{K^-_{n_1} (x_1 P^+ ) K^0_{n_2} (x_2 P^+)} V_{\text{strange}} \ket{\pi^-_n (P^+)} =
\frac{2 \pi}{P^+} \delta(x_1 +x_2 - 1) \varGamma_{n, n_1, n_2} (x_1, x_2),
%---------------------------
%---------------------------
\end{align}
%---------------------------
where $x_1$, $x_2$ denote the light-cone momentum fractions of $K^-_{n_1}$ and $K^0_{n_2}$ with respect to $\pi^-_n$, subject to the constraint $x_1+x_2=1$.
The triple-meson vertex function $\varGamma$  was first given by Callan, Coote and Gross~\cite{Callan:1975ps}.
In our case, it can be expressed in terms of the convolution of the LCWFs of $\pi_n$, $K^-_{n_1}$ and  $K^0_{n_2}$:
%---------------------------
\begin{align}
%---------------------------
\varGamma_{n,n_1,n_2}\left(x_1, x_2\right)
    =
   4 \lambda \sqrt{\frac{\pi}{N_{c}}}
    &\left[
        \int^1_{x_1}dy_1\int^{x_1}_0 dy_2\frac{1}{(y_2-y_1)^{2}}
        \varphi^{d\bar{u}}_{n}\left( 1-y_1\right)
        \varphi^{s\bar{u}}_{n_1}\left(1-\frac{y_2}{x_1}\right)
        \varphi^{d\bar{s}}_{n_2}\left(\frac{1-y_1}{x_2}\right)\right.
%---------------------------
\nn\\
%---------------------------
&\left.-\int^1_{x_2}dy_1\int^{x_2}_0 dy_2\frac{1}{(y_2-y_1)^{2}}
        \varphi^{d\bar{u}}_{n}\left( y_1\right)
        \varphi^{d\bar{s}}_{n_2}\left(\frac{y_2}{x_2}\right)
        \varphi^{s\bar{u}}_{n_1}\left(\frac{y_1-x_2}{x_1}\right) \right].
%---------------------------
\label{eq:GammaVrtx}
%---------------------------
\end{align}
%---------------------------
The corresponding Feynman diagrams are shown in Fig.~\ref{fig:three_meson_vertex}. We stress that this is a nonperturbative result, since all possible
planar gluon exchange diagrams have been resummed.

Substituting \eqref{eq:bilinear}, \eqref{NLO:Fock:component:pi:minus} and \eqref{LS:matrix:element} into \eqref{Def:strange:in:pion:PDF},
and repeatedly using \eqref{m_commute} to compute the vacuum matrix elements of the product of mesonic creation and annihilation operators,
we can work out the functional form of the $s$ and $\bar{s}$ PDF inside the $\pi^-_n$~\footnote{More technical details about this sort of
derivation can be found in our preceding work on intrinsic charm PDF of a flavor neutral meson~\cite{Hu:2022wsf}.}:
%---------------------------
\bsq
%---------------------------
\begin{align}
%---------------------------
f_{s/\pi^-_n}(x)\!= \!\! \sum_{n_1,n_2,n_3}
    \int_x^{1} dx_1&
    \frac{
        \varGamma_{n,n_1, n_2}(x_1, 1-x_1)
        \varGamma_{n,n_3, n_2}(x_1, 1-x_1)
    }{16\pi  x_{1}(1-x_1)}\left(
        \frac{\mu^{2}_{s \bar{u}, {n_1}}}{2x_1}\!+\!
        \frac{\mu^{2}_{d \bar{s}, {n_2}}}{2(1-x_1)}\!-\!
        \frac{\mu^{2}_{d \bar{u}, {n}}}{2}
    \right)^{-1}
%---------------------------
\non
%---------------------------
   & \times
    \left(
        \frac{\mu^{2}_{s \bar{u}, {n_3}}}{2x_1}\!+\!
        \frac{\mu^{2}_{d \bar{s}, {n_2}}}{2(1-x_1)}\!-\!
        \frac{\mu^{2}_{d \bar{u}, {n}}}{2}
    \right)^{-1} \!
    \left[
        \frac{1}{x_1}
        {
            \varphi^{s\bar{u}}_{n_1}\!\left(\frac{x}{x_1}\right)
            \varphi^{s\bar{u}}_{n_3}\!\left(\frac{x}{x_1}\right)
        }
    \right],
%---------------------------
\label{eq:icpdf}
%---------------------------
\\
%---------------------------
f_{\bar{s}/\pi^-_n}(x) \!= \!\! \sum_{n_1,n_2,n_3}
    \int_0^{1-x} \!\!\!dx_1 &
    \frac{
        \varGamma_{n,n_1, n_2}(x_1, 1-x_1)
        \varGamma_{n,n_1, n_3}(x_1, 1-x_1)
    }{16\pi  x_{1}(1-x_1)}   \left(
        \frac{\mu^{2}_{s \bar{u}, {n_1}}}{2x_1}\!+\!
        \frac{\mu^{2}_{d \bar{s}, {n_2}}}{2(1-x_1)}\!-\!
        \frac{\mu^{2}_{d \bar{u}, {n}}}{2}
    \right)^{-1}
%---------------------------
\non
%---------------------------
   & \times
    \left(
        \frac{\mu^{2}_{s \bar{u}, {n_1}}}{2x_1}\!+\!
        \frac{\mu^{2}_{d \bar{s}, {n_3}}}{2(1-x_1)}\!-\!
        \frac{\mu^{2}_{d \bar{u}, {n}}}{2}
    \right)^{-1} \!
    \left[
        \frac{1}{1-x_1}
       {
            \varphi^{d\bar{s}}_{n_2}\!\left(1\!-\!\frac{x}{1\!-\!x_1}\right)
            \varphi^{d\bar{s}}_{n_3}\!\left(1\!-\!\frac{x}{1\!-\!x_1}\right)
        }
    \right],
%---------------------------
\label{eq:acpdf}
\end{align}
%---------------------------
\label{s:sbar:PDF:main:formula}
%---------------------------
\esq
%---------------------------
The analytical forms of $f_{s/\pi^-}(x)$ and $f_{\bar{s}/\pi^-}(x)$ do differ from each other.

Equation~\eqref{s:sbar:PDF:main:formula} represents the major new result of this work,
which represents the rigorous expressions of the strange/antistrange PDF in a light flavored meson ($\pi^-$)
in 't Hooft model.
A schematic diagram to illustrate this formula is shown in the left panel of Fig.~\ref{fig:factrize}.
Infinite towers of the excited $K^-$ and ${K^0}$ mesons have to be included in the sum.
Note the principal quantum numbers $n_1$, $n_2$, $n_3$ are summed independently.
We will discuss the connection between this rigorous result and the prediction made in MCM.

\section{Strange-antistrange PDFs of $\pi^-$ in MCM}\label{sec:MCM}

\begin{figure}[htbp]
        \centering
        \includegraphics[scale=1.2]{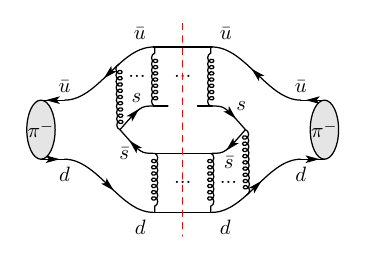}
        \includegraphics[scale=1.2]{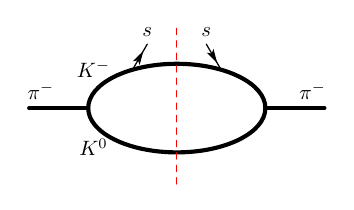}
    \caption{Schematic Feynman diagrams to illustrate the strange and antistrange PDFs of $\pi^-$,
    deduced from the first principle (left) and from meson cloud model.}
    \label{fig:factrize}
\end{figure}

Meson cloud model\cite{Sullivan:1971kd} provides a convenient and intuitive platform to estimate the intrinsic strange content inside the nucleon,
assuming that the nucleon has a non-negligible pentaquark component characterized by a stranged meson and a stranged baryon due to inevitable
quantum fluctuation~\cite{Koepf:1994an,Paiva:1996dd,Hobbs:2013bia,Melnitchouk:1997ig}. In line with the gist of the MCM, the strange content of $\pi^-$ in ${\rm QCD}_2$ originates from
the non-negligible probability for the $\pi^-$ to quantum fluctuate into the infinite towers of $K^-$ and $K^0$ pair.
Specifically, the strange quark PDF of $\pi^-_n$ can be expressed in terms of the convolution between
the transition probability $\pi^-_n \to K^-_{n_1} + K^0_{n_2}$,  $\mathcal{F}_{n,n_1,n_2}(x)$, and
the strange quark PDF of the $K^-(s\bar{u})$ mesons~\cite{Hu:2022wsf}:
%---------------------------
\beq
%---------------------------
	f_{s/\pi^-_n}(x) =
	\sum_{n_1,n_2}
	\int^1_x \frac{dy}{y}
	\mathcal{F}_{n,n_1,n_2} (y) f_{s/K_{n_1}^-}\left({x\over y}\right).
%---------------------------
\label{Recipe:MCM:Intrinisc:Strange}
%---------------------------
\eeq
%---------------------------
Similarly, one can obtain the $\overline{s}$ PDF provided that the strange quark PDF of the $K^-(s\bar{u})$
is replaced with the antistrange PDF of the $K^0(d\bar{s})$ meson.
A schematic Feynman diagram to picturise the MCM is shown in the right panel of Fig.~\ref{fig:factrize}.

The transition probability is linked with the triple-meson vertex via~\cite{Hu:2022wsf}
%---------------------------
\begin{align}
%---------------------------
	\mathcal{F}_{n,n_1n_2}(x_1)  =
	\frac{
		[\varGamma_{n,n_1, n_2}(x_1,1-x_1)]^2
	}{16\pi  x_{1}(1-x_1)}\!  \left(
	\frac{\mu^{2}_{K^-_{n_1}}}{2x_1}+
	\frac{\mu^{2}_{{K}^0_{n_2}}}{2(1-x_1)}-
	\frac{\mu^{2}_{\pi^-_{n}}}{2}
	\right)^{\!\!-2}.\label{eq:trns_prob}
	%---------------------------
\end{align}
%---------------------------

The strange PDF of the $K^-$ meson in  \eqref{Recipe:MCM:Intrinisc:Strange} is simply the square of the LCWF of the $K^-$.
After some manipulation, we obtain the MCM predictions to
the strange/antistrange PDFs of $\pi^-_n$ in 't Hooft model:
%---------------------------
\bsq
%---------------------------
\begin{align}
%---------------------------
f^{\rm MCM}_{s/\pi^-_n}(x) &\!= \!\! \sum_{n_1,n_2}
    \int_x^1 dx_1
    \frac{
        [\varGamma_{n,n_1, n_2}(x_1, 1-x_1)]^2
    }{16\pi  x_{1}(1-x_1)}\left(
        \frac{\mu^{2}_{s \bar{u}, {n_1}}}{2x_1}\!+\!
        \frac{\mu^{2}_{d \bar{s}, {n_2}}}{2(1-x_1)}\!-\!
        \frac{\mu^{2}_{d \bar{u}, {n}}}{2}
    \right)^{-2}
    \left[
        \frac{1}{x_1}
        \left(
        {
            \varphi^{s\bar{u}}_{n_1}\!\left(\frac{x}{x_1}\right)
        } \right)^2
    \right],
%---------------------------
\label{eq:mcm icpdf}
%---------------------------
\\
%---------------------------
f^{\rm MCM}_{\overline{s}/\pi^-_n}(x) & \!= \!\! \sum_{n_1,n_2}
    \int_0^{1-x} dx_1
    \frac{
        [\varGamma_{n,n_1, n_2}(x_1, 1-x_1)]^2
    }{16\pi  x_{1}(1-x_1)}   \left(
        \frac{\mu^{2}_{s \bar{u}, {n_1}}}{2x_1}\!+\!
        \frac{\mu^{2}_{d \bar{s}, {n_2}}}{2(1-x_1)}\!-\!
        \frac{\mu^{2}_{d \bar{u}, {n}}}{2}
    \right)^{-2}
    \left[
        \frac{1}{1-x_1}
        \left(
       {
       \varphi^{d\bar{s}}_{n_2}\!\left(1\!-\!\frac{x}{1\!-\!x_1}\right)
        }\right)^2
    \right],
%---------------------------
\label{eq:mcm acpdf}
%---------------------------
\end{align}
%---------------------------
\label{eq:mcm icacpdf}
%---------------------------
\esq
%---------------------------
Interestingly, these MCM predictions can be obtained from the rigourous results \eqref{s:sbar:PDF:main:formula} by keeping only the
diagonal terms in the sum~\footnote{In the phenomenological practice in real world, the term MCM is often interchangeably used with
the {\it naive} MCM, where only the lowest-lying strange mesons are kept in the sum.}.
Concretely speaking, the MCM prediction for strange PDF can be obtained
by setting $n_3=n_1$ in \eqref{eq:icpdf}, while the MCM prediction for antistrange PDF can be obtained
by setting $n_3=n_2$ in the sum in \eqref{eq:acpdf}.

\section{Numerical Results}
\label{sec:numerical}

As is evident in \eqref{s:sbar:PDF:main:formula} and \eqref{eq:mcm icacpdf},
the key ingredients of the intrinsic strange PDF
are various LCWFs of the $\pi^-$ and the excited $K^-$ and $K^0$ states.
The precise knowledge about these LCWFs are mandatory to make a trustworthy numerical predictions.

We follow the numerical recipe outlined in Ref.~\cite{Jia:2017uul} to numerically solve the 't Hooft equation \eqref{tHooft:equation}
with high precision. To handle with the hadrons made of the light quarks, we adopt the following
parametrization for the mesonic LCWFs~\cite{tHooft:1974pnl}
%---------------------------
\begin{align}
%---------------------------
\label{eq:wave function}
%---------------------------
\varphi^{u/d}_n(x) = c^{u/d}_{0} x^{\beta^{u/d}} (1-x)^{2-\beta^{u/d}} + c^{u/d}_{1} x^{2-\beta^{d/u}}(1-x)^{\beta^{d/u}}+ \sum_{n=1}^{N} c^{u/d}_n \sin(n\pi x),
%---------------------------
\end{align}
%---------------------------
where $\beta^{u/d} (0 < \beta^{u/d} <  \pi / 2)$ are determined by the boundary condition
$ \pi \beta^{u/d} \cot \pi \beta^{u/d} = 1 -  m^2_{u/d} / 2 \lambda$.

\subsection{Quark mass setting in \QCDtwo}
\label{sec:quark:mass:setting}

\begin{table}[htb!]
    \renewcommand{\arraystretch}{1.4}
    \setlength{\tabcolsep}{0.0075\textwidth}
\begin{center}
%{$n=1$}\\
\vspace{0.5em}
\begin{tabular}{ccccccccc}
\hline Meson & $m_q~/\massu$  & $\mu_0/\massu$ & $\mu_1/\massu$ & $\mu_2/\massu$ & $\mu_3/\massu$ & $\mu_4/\massu$ & $\mu_5/\massu$  \\
\hline\hline
\text{$ \pi^-$} & {\makecell{$m_u = 0.0285$ \\ $m_d = 0.0570$}}  & $0.402$ & $2.489$ & $3.816$ & $4.850$ & $5.723$  & $6.491$    \\
\hline
\text{$ K^-$} &  {\makecell{$m_u = 0.0285$ \\ $m_s = 0.791$}}  & $1.452$ & $3.155$  & $4.357$ & $5.321$ & $6.148$  & $6.882$    \\
\hline
\text{$ K^0$} &  {\makecell{$m_d = 0.057$ \\ $m_s = 0.791$}}  & $1.484$ & $3.178$  & $4.374$ & $5.335$ & $6.160$  & $6.893$    \\
\hline
\end{tabular}
\caption{The masses of the $u$, $d$, $s$, as well as the mass spectra of the first six low-lying states in the $\pi^-$, $K^-$ and $K^0$ families.}
\label{tab:meson mass}
\end{center}
\end{table}

The 't Hooft coupling is chosen as $\sqrt{2\lambda} = 340\;\mathrm{MeV}$, in conformity with
the string tension in realistic four-dimensional QCD~\cite{Burkardt:2000uu}.

To fathom the nonvanishing $s$-$\bar{s}$ asymmetry inside a pion, we must implement the isospin breaking effect, {\it viz.}, consider the case with
unequal $u$ and $d$ masses. We assume $m_u/m_d=1/2$, the same as in realistic ${\rm QCD}_4$ determined by the chiral perturbation theory.
Requiring $m_{\pi^-} = 137\;\mathrm{MeV}= 0.402\,\sqrt{2\lambda}$, we obtain $m_u = 0.0285\,\sqrt{2\lambda}$ by
solving 't Hooft equation. With such light $u$ and $d$ quark masses, we have checked that the Gell-Mann-Oakes-Renner relation is decently
satisfied for the ground-state pion.

We proceed to set the strange quark mass. Taking $m_{K^-} =494\;\mathrm{MeV}= 1.451\sqrt{2\lambda}$ as input, we determine
$m_s=0.791\sqrt{2\lambda}$ upon solving 't Hooft equation.
The quark masses together with the first few low-lying states in $\pi^-$, $K^-$, and $K^0$ families are enumerated
in Table~\ref{tab:meson mass}.

\subsection{Strange-antistrange asymmetry in the first excited $\pi^-$}

We follow \eqref{s:sbar:PDF:main:formula} and \eqref{eq:mcm icacpdf} to calculate intrinsic strange and antistrange PDFs in $\pi^-$ in ${\rm QCD}_2$,
deduced from both the first principle and the MCM, respectively.
It is found that retaining the first 60 excited states in the sum in \eqref{s:sbar:PDF:main:formula} and \eqref{eq:mcm icacpdf} already
exhibit satisfactory convergence behavior.
To quantify the extent of the $s$-$\bar{s}$ asymmetry, we define the following dimensionless ratio:
%---------------------------
\beq
%---------------------------
A^{s\bar{s}}_{\pi^-} \equiv {\delta s(x)\over \langle s \rangle} \equiv {f_{s/\msn}(x) - f_{\overline{s}/\msn} (x)\over \int_0^1 \! dx\, f_{s/\msn}(x)}.
%---------------------------
\label{eq:def delta c}
%---------------------------
\eeq
%---------------------------
Since the net strangeness in $\pi^-$ must vanish, so that $\langle s \rangle=\langle \bar{s} \rangle$ and
$\displaystyle \int^1_0 \!dx\, \delta s(x)=0$.

\begin{figure}[htb!]
        \centering
        \includegraphics[width=0.32\textwidth]{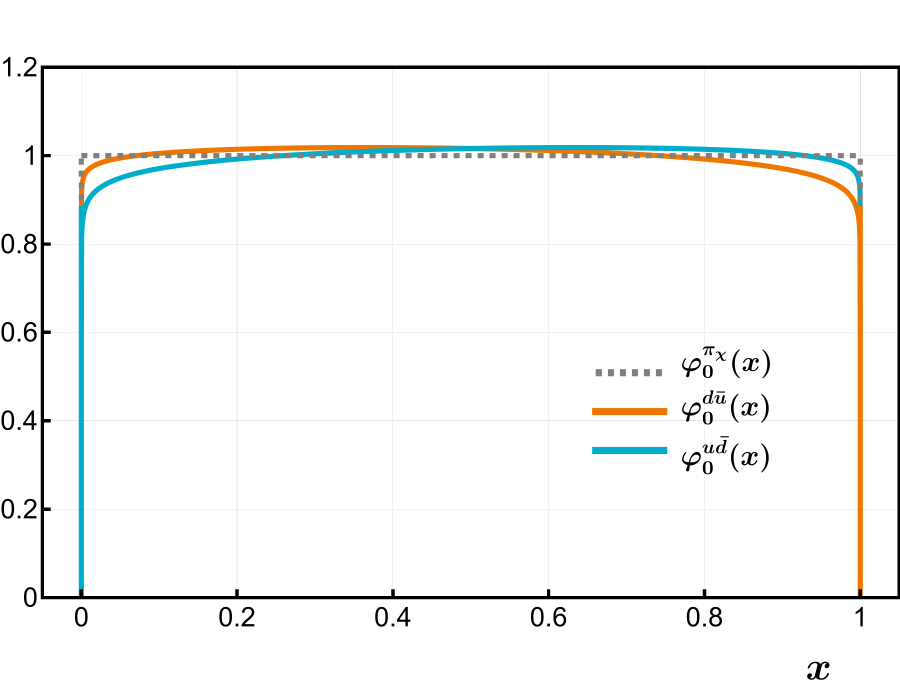}
        \includegraphics[width=0.32\textwidth]{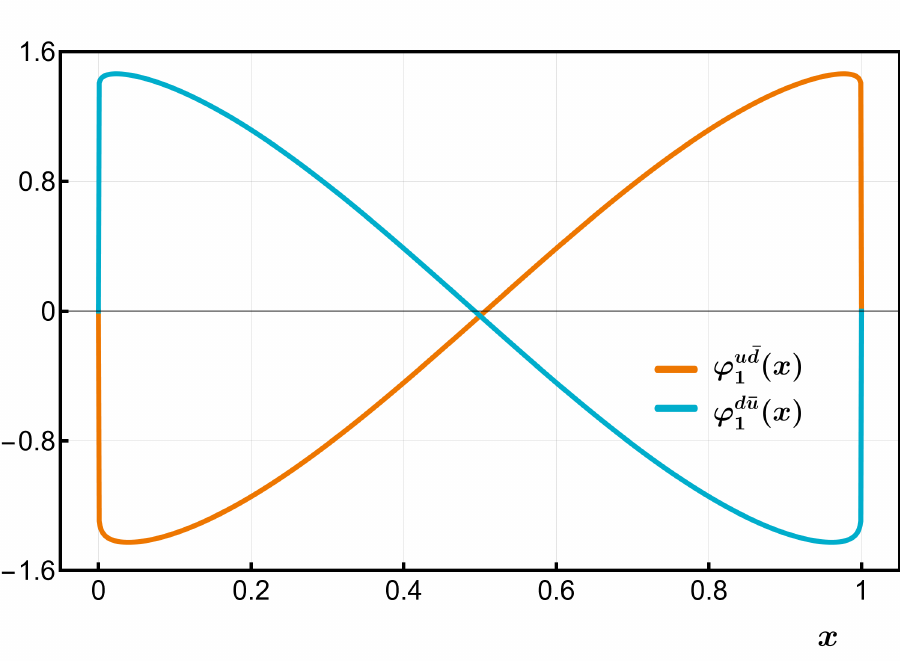}
        \includegraphics[width=0.32\textwidth]{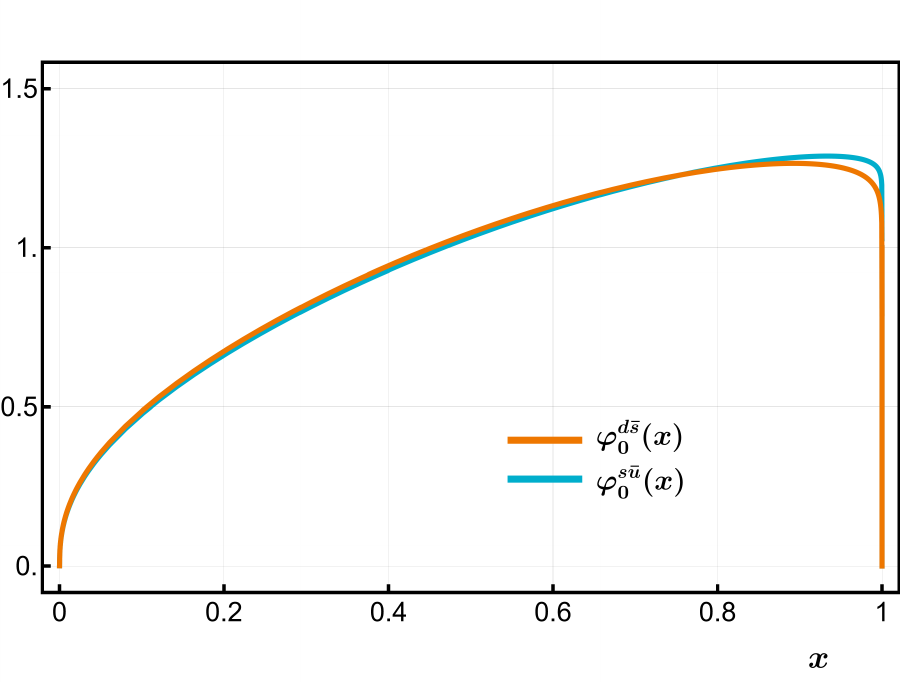}
    \caption{Profiles of LCWFs of various mesons, exemplified by the chiral and physical $\pi^-$ (left), the first excited $\pi^-$ (middle), and
    the lowest-lying $K^-$ and $K^0$ (right). }
\label{fig:dubarchiralpi}
\end{figure}

\begin{figure}[htb!]
        \centering
        \includegraphics[width=0.45\textwidth]{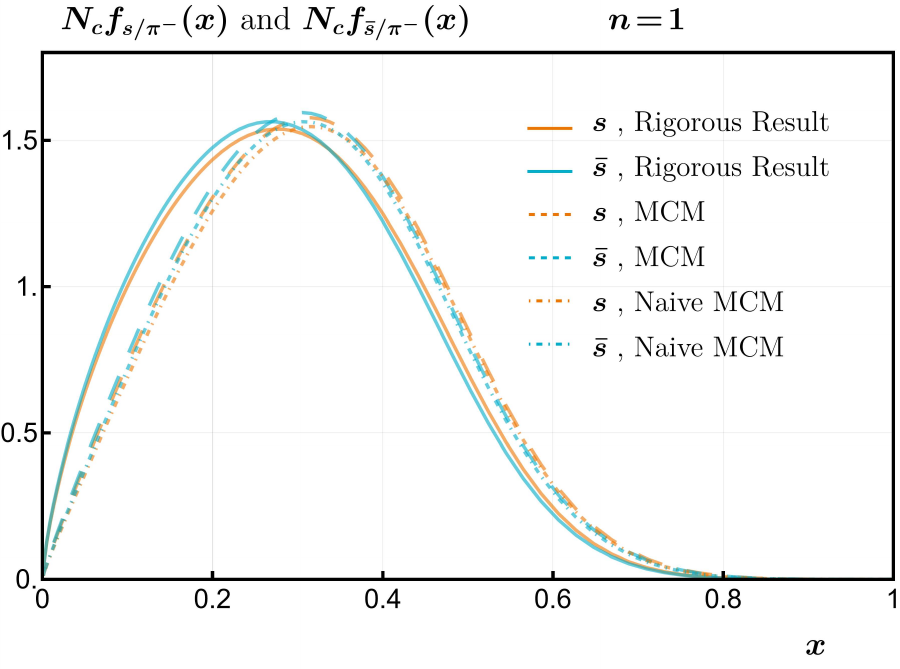}
        \includegraphics[width=0.45\textwidth]{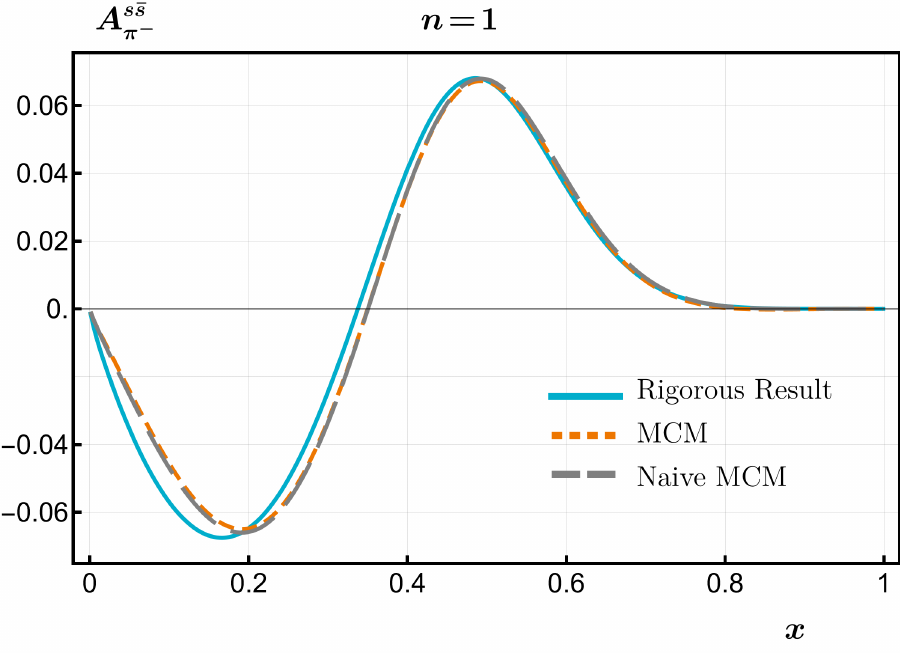}
    \caption{Strange and antistrange PDFs of the first excited $\pi^-$ (left), together with the corresponding $s$-$\bar{s}$ asymmetry (right).
For the sake of comparison, we juxtapose the predictions from the first principle of QCD and from the MCMs. }
    \label{fig:new_s deltac}
\end{figure}

The intrinsic strange distribution in the ground-state $\pi^-$ turns out to be extremely small and very challenging to accurately calculate.
The reason is somewhat accidental, mainly attributed to a peculiar feature of the 't Hooft model, that the triple meson interaction strength strictly vanishes when
one of the mesons is the chiral pion $\pi_\chi$, irrespective the magnitude of its momentum~\cite{Kalashnikova:2000dw}.
This can be verified by substituting the LCWF of the chiral pion, $\varphi_0^{\pi_\chi} (x)=\Theta(x)\Theta(1-x)$, into the triple meson vertex function
\eqref{eq:GammaVrtx}. With very light $u$ and $d$ quark masses as specified in  Table~\ref{tab:meson mass},
the profile of the LCWF of a ``physical" $\pi$
is quite close to that of the chiral $\pi$ (see the left panel of Fig.~\ref{fig:dubarchiralpi}),
so that the triple meson interaction strength gets severely suppressed,
which leads to exceedingly small strange PDF.

Therefore we choose to analyze the intrinsic strange and antistrange PDFs in the first excited $\pi^-$, whose LCWF is shown in the middle panel of Fig.~\ref{fig:dubarchiralpi}).

The seed of the $s$-$\bar{s}$ asymmetry in our case is the isospin breaking due to the $u$-$d$ mass difference.
As can be seen from the right panel of Fig.~\ref{fig:dubarchiralpi}),
though the LCWFs of ground-state $K^-$ and $K^0$ resemble with each other, there still exists some slight difference.
In the left panel of Fig.~\ref{fig:new_s deltac}, we juxtapose various $s$ and $\bar{s}$ PDFs in the first excited $\pi^-$,
predicted both from the first principle and two variants of MCM.
One observes the predictions from the first principle and the MCMs grossly agree, despite with some moderate difference.
The difference between rigorous results and MCM predictions is reflected in
the interference terms in \eqref{s:sbar:PDF:main:formula}, {\it e.g.}, those terms in the sum with $n_1\neq n_3$ or $n_2 \neq n_3$.
Curiously, we also notice that the results from the full MCM and naive MCM seem indistinguishable, which implies that the
sum over the highly excited strange meson does not yield a significant contribution.

In the right panel of Fig.~\ref{fig:new_s deltac}, we also plot the $s$-$\bar{s}$ asymmetry as a function of $x$.
The profiles predicted from the first principle and MCM bear a similar shape. It is interesting to see that
the asymmetry can reach per cents level. There is a sign change near $x\approx 0.4$.
There appear to have more $\bar{s}$ quarks in the low $x$ interval, and more $s$ quark in the high $x$ interval.
This pattern is likely strongly correlated with the fact $m_u<m_d$, as well as the parent hadron chosen as the first excited $\pi^-$.

\subsection{Charm-anticharm asymmetry in the first excited $\pi^-$}

It has been observed that the intrinsic charm content in a flavor-neutral 
light meson roughly scales as $1/m_c^6$ in ${\rm QCD}_2$~\cite{Hu:2022wsf},,
in stark contrast with the $1/m_c^2$ decrease in the realistic ${\rm QCD}_4$.
One may naturally wonder how the sea quark-antiquark asymmetry depends on the sea quark mass. 
In this subsection, we investigate the $c$-$\bar{c}$ asymmetry in the first excited $\pi^-$, 
simply repeating the preceding analysis with the strange quark replaced by the charm quark.

To match the mass of the lowest-lying charmonium in realistic world, we assume $m_c = 4.19 \sqrt{2\lambda}$~\cite{Jia:2017uul}.
We again specialize to the intrinsic charm and anticharm PDFs in the light flavored meson,  the first excited $\pi^-$.
We follow the same numerical recipe as expounded in \cite{Hu:2022wsf}, by imposing some convergence criteria 
about the infinite summation in \eqref{s:sbar:PDF:main:formula}.
A slower convergence pace is observed relative to the strange case.

\begin{figure}[htbp]
        \centering
        \includegraphics[width=0.85\textwidth]{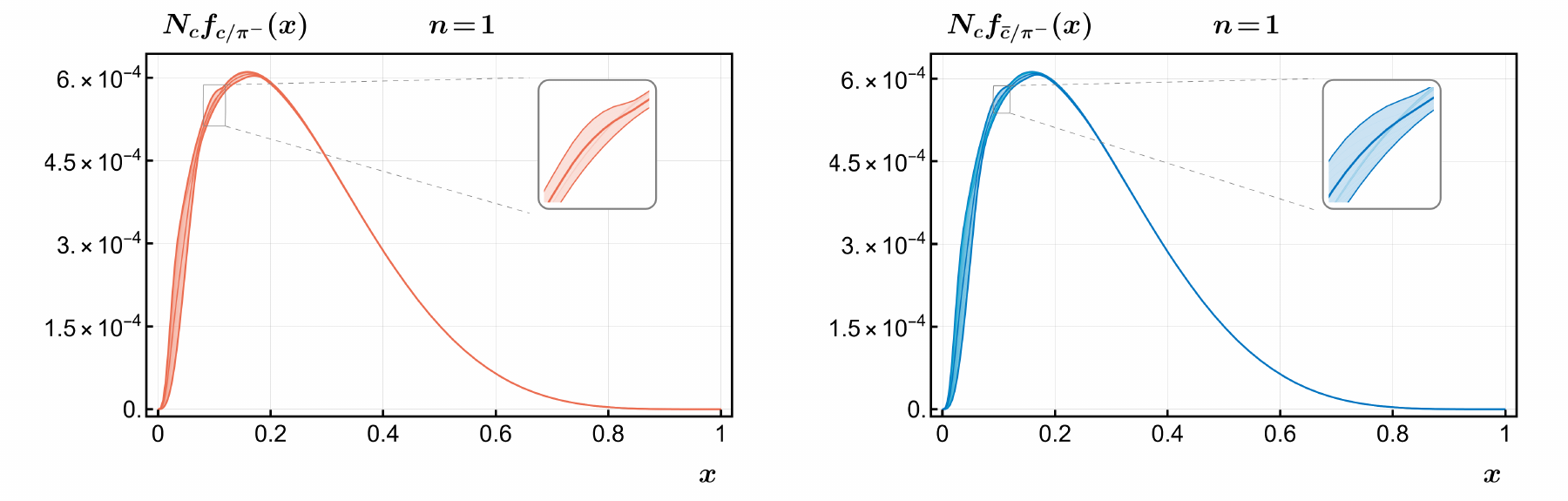}
      \caption{Intrinsic charm PDF (left) and anticharm PDF (right) in the first excited $\pi^-$. 
There is some technical nuisance in the small $x$ region, reflected by the oscillatory behavior due to truncation error.}
    \label{fig:icacpdf_subs}
\end{figure}

In Fig.~\ref{fig:icacpdf_subs}, we present the intrinsic $c$ and $\bar{c}$ PDFs of the first excited $\pi^-$. 
We adopt the same numerical strategy as in Ref.~\cite{Hu:2022wsf} to present the numerical results, {\it i.e.},
by treating the upper/lower envelopes of data as uncertainty band, while taking the average of the envelope as the 
central value.

\begin{figure}[htbp]
        \centering
        \includegraphics[width=0.32\textwidth]{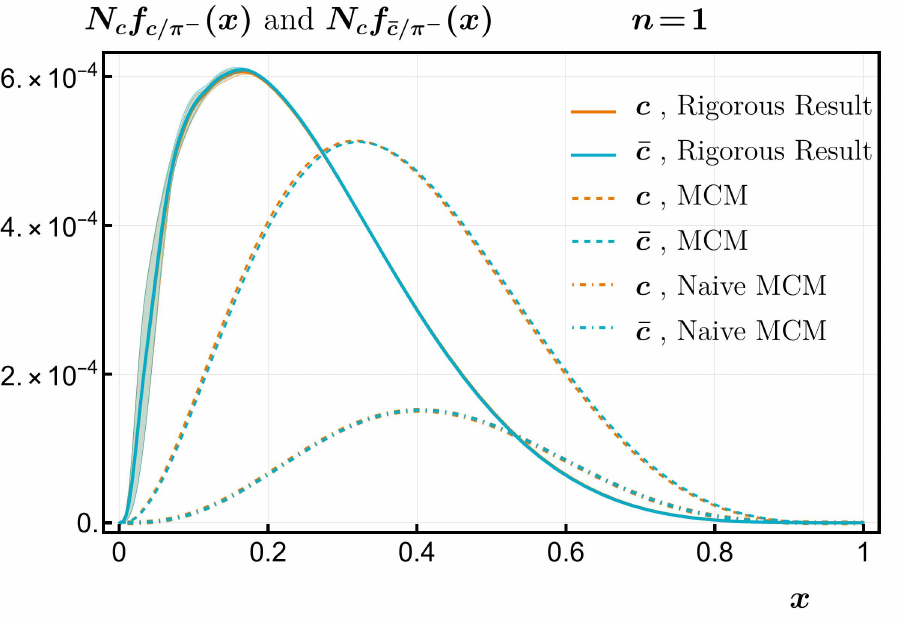}
        \includegraphics[width=0.32\textwidth]{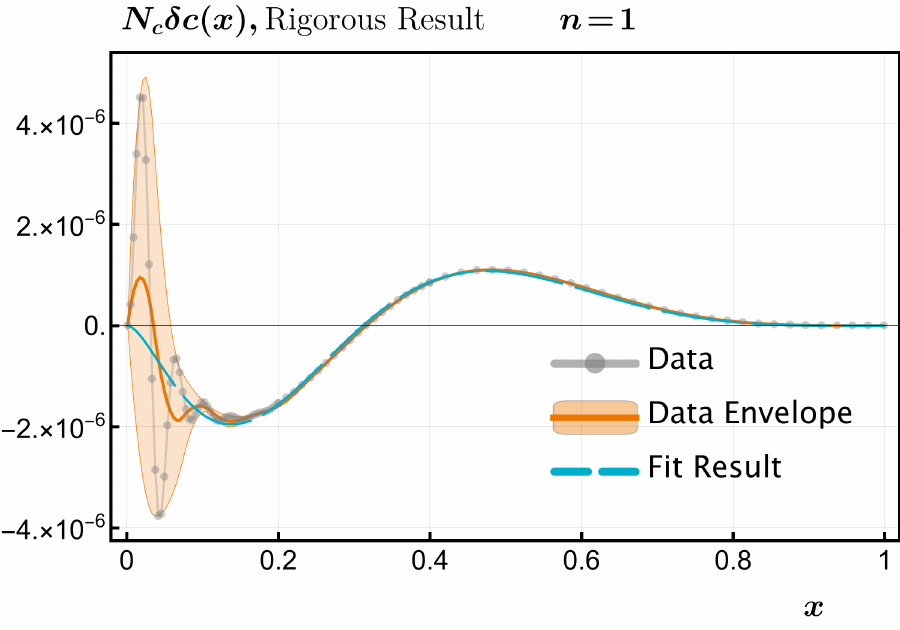}
        \includegraphics[width=0.32\textwidth]{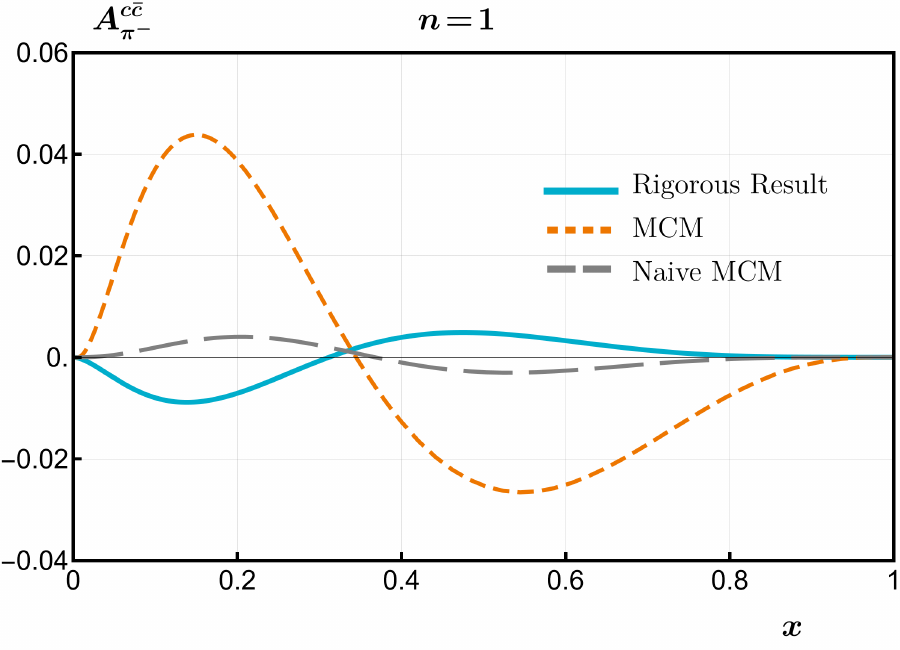}
\caption{The $c$ and $\bar{c}$ PDFs (left) and $c$-$\bar{c}$ asymmetry (right) for the first-excited state $\pi^-$.  
 The predictions from the rigorous approach and two variants of MCM are juxtaposed together.
The data processing for $\delta c$ is presented in the middle panel.}
\label{fig:icacpdf_2_msn}
\end{figure}

From the left panel of Fig.~\ref{fig:icacpdf_2_msn}, one clearly sees that the intrinsic charm content in the first excited $\pi^-$
become several orders of magnitude smaller than the intrinsic strange content. 
Curiously, as can be seen in the right panel of Fig.~\ref{fig:icacpdf_2_msn}, 
the magnitude of the $c$-$\bar{c}$ asymmetry can still reach the per cent level. 
In contrast to the case of the intrinsic strange,
the discrepancy between rigorous results and MCM predictions for intrinsic charm becomes pronounced.
In particular, we notice that the sign of $\delta c(x)$ have become opposite in the entire support of $x$.

\section{summary} \label{sec:summary}

In this work, we extend our preceding analysis of the intrinsic charm PDF in a flavor neutral light meson in 't Hooft model
to the strange-antistrange asymmetry in a light flavored meson such as charged pion.
The intrinsic strange content in $\pi^-$  has to stem from its higher Fock component,
which is composed of the infinite towers of excited $K^-$ and $K^0$ states,
thus characterizing an ${\cal O}(1/N_c)$ correction.
The analytical expressions for the $s$ and $\bar{s}$ PDFs are derived using the Hamiltonian approach within
light-front quantization, in terms of the convolution of mesonic light-cone wave functions.
For comparison, we also present the analytical forms of the strange and antistrange PDFs predicted by the meson cloud model.

In numerical analysis, we have attempted to mimic the realistic QCD by tuning the $u$, $d$ masses such that the physical pion
is recovered by fixing $m_u/m_d=1/2$.  Due to some peculiarity of chiral pion in two-dimensional QCD, the strange content of the ground-state
pion is exceedingly small, so we choose to investigate the $s$-$\bar{s}$ asymmetries inside the first excited $\pi^-$.
With $m_u/m_d=1/2$, the $s$-$\bar{s}$ and $c$-$\bar{c}$ asymmetries inside the first excited $\pi^-$ in
${\rm QCD}_{\rm 2}$ can reach per cents level. It is also found that the predicted $s$-$\bar{s}$ asymmetry from MCM
largely agree with that from the first principle calculation for $s$-$\bar{s}$ asymmetries.
But severe discrepancy is observed for the $c$-$\bar{c}$ asymmetry in the first excited $\pi^-$.

\section*{Acknowledgements}

This work was supported in part by the High Performance Computing Center of Central South University.
The work of S.~H., Y.~J. and Z.~M. is supported in part by the National Natural Science Foundation of China under Grants No. 11925506.
Z.~M. is also supported in part by the National Natural Science Foundation of China under Grants No. 12347145, No. 12022514.
The work of X.-N. X. and M.~L. Z. is supported by the National Natural Science Foundation of China under Grants No. 12275364.


\begin{thebibliography}{100}

%\cite{Martin:2009iq}
\bibitem{Martin:2009iq}
A.~D.~Martin, W.~J.~Stirling, R.~S.~Thorne and G.~Watt,
%``Parton distributions for the LHC,''
Eur. Phys. J. C \textbf{63}, 189-285 (2009)
doi:10.1140/epjc/s10052-009-1072-5
[arXiv:0901.0002 [hep-ph]].
%5624 citations counted in INSPIRE as of 30 Dec 2024

%\cite{Hou:2019efy}
\bibitem{Hou:2019efy}
T.~J.~Hou, J.~Gao, T.~J.~Hobbs, K.~Xie, S.~Dulat, M.~Guzzi, J.~Huston, P.~Nadolsky, J.~Pumplin and C.~Schmidt, \textit{et al.}
%``New CTEQ global analysis of quantum chromodynamics with high-precision data from the LHC,''
Phys. Rev. D \textbf{103}, no.1, 014013 (2021)
doi:10.1103/PhysRevD.103.014013
[arXiv:1912.10053 [hep-ph]].
%697 citations counted in INSPIRE as of 30 Dec 2024

%\cite{Ablat:2024muy}
\bibitem{Ablat:2024muy}
A.~Ablat, A.~Courtoy, S.~Dulat, M.~Guzzi, T.~J.~Hobbs, T.~J.~Hou, J.~Huston, K.~Mohan, H.~W.~Lin and P.~Nadolsky, \textit{et al.}
%``New results in the CTEQ-TEA global analysis of parton distributions in the nucleon,''
Eur. Phys. J. Plus \textbf{139}, no.12, 1063 (2024)
doi:10.1140/epjp/s13360-024-05865-x
[arXiv:2408.04020 [hep-ph]].
%4 citations counted in INSPIRE as of 23 Dec 2024

%\cite{NNPDF:2021njg}
\bibitem{NNPDF:2021njg}
R.~D.~Ball \textit{et al.} [NNPDF],
%``The path to proton structure at 1\% accuracy,''
Eur. Phys. J. C \textbf{82}, no.5, 428 (2022)
doi:10.1140/epjc/s10052-022-10328-7
[arXiv:2109.02653 [hep-ph]].
%418 citations counted in INSPIRE as of 30 Dec 2024

%\cite{NNPDF:2024dpb}
\bibitem{NNPDF:2024dpb}
R.~D.~Ball \textit{et al.} [NNPDF],
%``Determination of the theory uncertainties from missing higher orders on NNLO parton distributions with percent accuracy,''
Eur. Phys. J. C \textbf{84}, no.5, 517 (2024)
doi:10.1140/epjc/s10052-024-12772-z
[arXiv:2401.10319 [hep-ph]].
%28 citations counted in INSPIRE as of 23 Dec 2024

%\cite{Hobbs:2019gob}
\bibitem{Hobbs:2019gob}
T.~J.~Hobbs, B.~T.~Wang, P.~M.~Nadolsky and F.~I.~Olness,
%``Charting the coming synergy between lattice QCD and high-energy phenomenology,''
Phys. Rev. D \textbf{100}, no.9, 094040 (2019)
doi:10.1103/PhysRevD.100.094040
[arXiv:1904.00022 [hep-ph]].
%34 citations counted in INSPIRE as of 19 Dec 2024

%\cite{Constantinou:2020hdm}
\bibitem{Constantinou:2020hdm}
M.~Constantinou, A.~Courtoy, M.~A.~Ebert, M.~Engelhardt, T.~Giani, T.~Hobbs, T.~J.~Hou, A.~Kusina, K.~Kutak and J.~Liang, \textit{et al.}
%``Parton distributions and lattice-QCD calculations: Toward 3D structure,''
Prog. Part. Nucl. Phys. \textbf{121}, 103908 (2021)
doi:10.1016/j.ppnp.2021.103908
[arXiv:2006.08636 [hep-ph]].
%190 citations counted in INSPIRE as of 23 Dec 2024

%\cite{Ji:2020ect}
\bibitem{Ji:2020ect}
X.~Ji, Y.~S.~Liu, Y.~Liu, J.~H.~Zhang and Y.~Zhao,
%``Large-momentum effective theory,''
Rev. Mod. Phys. \textbf{93}, no.3, 035005 (2021)
doi:10.1103/RevModPhys.93.035005
[arXiv:2004.03543 [hep-ph]].
%269 citations counted in INSPIRE as of 23 Dec 2024

%\cite{NuTeV:2001whx}
\bibitem{NuTeV:2001whx}
G.~P.~Zeller \textit{et al.} [NuTeV],
%``A Precise Determination of Electroweak Parameters in Neutrino Nucleon Scattering,''
Phys. Rev. Lett. \textbf{88}, 091802 (2002)
[erratum: Phys. Rev. Lett. \textbf{90}, 239902 (2003)]
doi:10.1103/PhysRevLett.88.091802
[arXiv:hep-ex/0110059 [hep-ex]].
%862 citations counted in INSPIRE as of 24 Dec 2024

%\cite{NuTeV:2001dfo}
\bibitem{NuTeV:2001dfo}
M.~Goncharov \textit{et al.} [NuTeV],
%``Precise Measurement of Dimuon Production Cross-Sections in $\nu_{\mu}$ Fe and $\bar{\nu}_{\mu}$ Fe Deep Inelastic Scattering at the Tevatron.,''
Phys. Rev. D \textbf{64}, 112006 (2001)
doi:10.1103/PhysRevD.64.112006
[arXiv:hep-ex/0102049 [hep-ex]].
%397 citations counted in INSPIRE as of 19 Dec 2024

%\cite{NuTeV:2002ryj}
\bibitem{NuTeV:2002ryj}
G.~P.~Zeller \textit{et al.} [NuTeV],
%``On the Effect of Asymmetric Strange Seas and Isospin Violating Parton Distribution Functions on $\sin^{2} \theta_{W}$ Measured in the NuTeV Experiment,''
Phys. Rev. D \textbf{65}, 111103 (2002)
[erratum: Phys. Rev. D \textbf{67}, 119902 (2003)]
doi:10.1103/PhysRevD.65.111103
[arXiv:hep-ex/0203004 [hep-ex]].
%153 citations counted in INSPIRE as of 19 Dec 2024

%\cite{Brodsky:1996hc}
\bibitem{Brodsky:1996hc}
S.~J.~Brodsky and B.~Q.~Ma,
%``The Quark / anti-quark asymmetry of the nucleon sea,''
Phys. Lett. B \textbf{381}, 317-324 (1996)
doi:10.1016/0370-2693(96)00597-7
[arXiv:hep-ph/9604393 [hep-ph]].
%285 citations counted in INSPIRE as of 13 Dec 2024

%\cite{Wang:2016ndh}
\bibitem{Wang:2016ndh}
X.~G.~Wang, C.~R.~Ji, W.~Melnitchouk, Y.~Salamu, A.~W.~Thomas and P.~Wang,
%``Strange quark asymmetry in the proton in chiral effective theory,''
Phys. Rev. D \textbf{94}, no.9, 094035 (2016)
doi:10.1103/PhysRevD.94.094035
[arXiv:1610.03333 [hep-ph]].
%32 citations counted in INSPIRE as of 11 Dec 2024

%\cite{Du:2017nzy}
\bibitem{Du:2017nzy}
X.~Du and B.~Q.~Ma,
%``Strange quark-antiquark asymmetry of nucleon sea from $\Lambda/\bar\Lambda$ polarization,''
Phys. Rev. D \textbf{95}, no.1, 014029 (2017)
doi:10.1103/PhysRevD.95.014029
[arXiv:1701.04945 [hep-ph]].
%7 citations counted in INSPIRE as of 14 Nov 2024

%\cite{Salajegheh:2015xoa}
\bibitem{Salajegheh:2015xoa}
M.~Salajegheh,
%``Intrinsic strange distributions in the nucleon from the light-cone models,''
Phys. Rev. D \textbf{92}, no.7, 074033 (2015)
doi:10.1103/PhysRevD.92.074033
[arXiv:1602.00154 [hep-ph]].
%12 citations counted in INSPIRE as of 09 Dec 2024

%\cite{Vega:2015hti}
\bibitem{Vega:2015hti}
A.~Vega, I.~Schmidt, T.~Gutsche and V.~E.~Lyubovitskij,
%``Nonperturbative contribution to the strange-antistrange asymmetry of the nucleon sea,''
Phys. Rev. D \textbf{93}, no.5, 056001 (2016)
doi:10.1103/PhysRevD.93.056001
[arXiv:1511.06476 [hep-ph]].
%16 citations counted in INSPIRE as of 25 Sep 2024

%\cite{Catani:2004nc}
\bibitem{Catani:2004nc}
S.~Catani, D.~de Florian, G.~Rodrigo and W.~Vogelsang,
%``Perturbative generation of a strange-quark asymmetry in the nucleon,''
Phys. Rev. Lett. \textbf{93}, 152003 (2004)
doi:10.1103/PhysRevLett.93.152003
[arXiv:hep-ph/0404240 [hep-ph]].
%107 citations counted in INSPIRE as of 25 Sep 2024

%\cite{Sullivan:1971kd}
\bibitem{Sullivan:1971kd}
J.~D.~Sullivan,
%``One pion exchange and deep inelastic electron - nucleon scattering,''
Phys. Rev. D \textbf{5}, 1732-1737 (1972)
doi:10.1103/PhysRevD.5.1732
%362 citations counted in INSPIRE as of 28 Nov 2024

%\cite{Brodsky:1980pb}
\bibitem{Brodsky:1980pb}
S.~J.~Brodsky, P.~Hoyer, C.~Peterson and N.~Sakai,
%``The Intrinsic Charm of the Proton,''
Phys. Lett. B \textbf{93}, 451-455 (1980)
doi:10.1016/0370-2693(80)90364-0
%886 citations counted in INSPIRE as of 17 Dec 2024

%\cite{Brodsky:1981se}
\bibitem{Brodsky:1981se}
S.~J.~Brodsky, C.~Peterson and N.~Sakai,
%``Intrinsic Heavy Quark States,''
Phys. Rev. D \textbf{23}, 2745 (1981)
doi:10.1103/PhysRevD.23.2745
%579 citations counted in INSPIRE as of 11 Dec 2024

%\cite{Franz:2000ee}
\bibitem{Franz:2000ee}
M.~Franz, M.~V.~Polyakov and K.~Goeke,
%``Heavy quark mass expansion and intrinsic charm in light hadrons,''
Phys. Rev. D \textbf{62}, 074024 (2000)
doi:10.1103/PhysRevD.62.074024
[arXiv:hep-ph/0002240 [hep-ph]].
%160 citations counted in INSPIRE as of 25 Sep 2024

%\cite{LHCb:2021stx}
\bibitem{LHCb:2021stx}
R.~Aaij \textit{et al.} [LHCb],
%``Study of Z Bosons Produced in Association with Charm in the Forward Region,''
Phys. Rev. Lett. \textbf{128}, no.8, 082001 (2022)
doi:10.1103/PhysRevLett.128.082001
[arXiv:2109.08084 [hep-ex]].
%53 citations counted in INSPIRE as of 13 Dec 2024

%\cite{Ball:2022qks}
\bibitem{Ball:2022qks}
R.~D.~Ball \textit{et al.} [NNPDF],
%``Evidence for intrinsic charm quarks in the proton,''
Nature \textbf{608}, no.7923, 483-487 (2022)
doi:10.1038/s41586-022-04998-2
[arXiv:2208.08372 [hep-ph]].
%77 citations counted in INSPIRE as of 23 Dec 2024

%\cite{NNPDF:2023tyk}
\bibitem{NNPDF:2023tyk}
R.~D.~Ball \textit{et al.} [NNPDF],
%``Intrinsic charm quark valence distribution of the proton,''
Phys. Rev. D \textbf{109}, no.9, L091501 (2024)
doi:10.1103/PhysRevD.109.L091501
[arXiv:2311.00743 [hep-ph]].
%25 citations counted in INSPIRE as of 17 Dec 2024

%\cite{tHooft:1974pnl}
\bibitem{tHooft:1974pnl}
G.~'t Hooft,
%``A Two-Dimensional Model for Mesons,''
Nucl. Phys. B \textbf{75}, 461-470 (1974)
doi:10.1016/0550-3213(74)90088-1
%2133 citations counted in INSPIRE as of 26 Dec 2024

%\cite{Callan:1975ps}
\bibitem{Callan:1975ps}
C.~G.~Callan, Jr., N.~Coote and D.~J.~Gross,
%``Two-Dimensional Yang-Mills Theory: A Model of Quark Confinement,''
Phys. Rev. D \textbf{13}, 1649 (1976)
doi:10.1103/PhysRevD.13.1649
%374 citations counted in INSPIRE as of 10 Dec 2024

%\cite{Einhorn:1976uz}
\bibitem{Einhorn:1976uz}
M.~B.~Einhorn,
%``Form-Factors and Deep Inelastic Scattering in Two-Dimensional Quantum Chromodynamics,''
Phys. Rev. D \textbf{14}, 3451 (1976)
doi:10.1103/PhysRevD.14.3451
%258 citations counted in INSPIRE as of 05 Oct 2024

%\cite{Einhorn:1976ax}
\bibitem{Einhorn:1976ax}
M.~B.~Einhorn, S.~Nussinov and E.~Rabinovici,
%``Meson Scattering in Quantum Chromodynamics in Two-Dimensions,''
Phys. Rev. D \textbf{15}, 2282 (1977)
doi:10.1103/PhysRevD.15.2282
%42 citations counted in INSPIRE as of 25 Sep 2024

%\cite{Brower:1977hx}
\bibitem{Brower:1977hx}
R.~C.~Brower, J.~R.~Ellis, M.~G.~Schmidt and J.~H.~Weis,
%``Hadron Scattering in Two-Dimensional QCD. 1. Formalism and Leading Order Calculations,''
Nucl. Phys. B \textbf{128}, 131-174 (1977)
doi:10.1016/0550-3213(77)90303-0
%50 citations counted in INSPIRE as of 25 Sep 2024

%\cite{Brower:1977as}
\bibitem{Brower:1977as}
R.~C.~Brower, J.~R.~Ellis, M.~G.~Schmidt and J.~H.~Weis,
%``Hadron Scattering in Two-Dimensional QCD. 2. Second Order Calculations, Multi-Regge and Inclusive Reactions,''
Nucl. Phys. B \textbf{128}, 175-203 (1977)
doi:10.1016/0550-3213(77)90304-2
%64 citations counted in INSPIRE as of 25 Sep 2024

%\cite{Brower:1978wm}
\bibitem{Brower:1978wm}
R.~C.~Brower, W.~L.~Spence and J.~H.~Weis,
%``Bound States and Asymptotic Limits for {QCD} in Two-dimensions,''
Phys. Rev. D \textbf{19}, 3024 (1979)
doi:10.1103/PhysRevD.19.3024
%43 citations counted in INSPIRE as of 01 Oct 2024

%\cite{Hu:2022wsf}
\bibitem{Hu:2022wsf}
S.~Hu, Y.~Jia, Z.~Mo, X.~Xiong and M.~Zhu,
%``Parton distributions of intrinsic charm in two-dimensional QCD,''
Phys. Rev. D \textbf{108}, no.9, 094040 (2023)
doi:10.1103/PhysRevD.108.094040
[arXiv:2211.16489 [hep-ph]].
%2 citations counted in INSPIRE as of 04 Dec 2024

%\cite{Kikkawa:1980dc}
\bibitem{Kikkawa:1980dc}
K.~Kikkawa,
%``A GAUGE INVARIANT APPROACH TO TWO-DIMENSIONAL QUANTUM CHROMODYNAMICS,''
Annals Phys. \textbf{135}, 222 (1981)
doi:10.1016/0003-4916(81)90154-8
%15 citations counted in INSPIRE as of 01 Oct 2024

%\cite{Nakamura:1981zi}
\bibitem{Nakamura:1981zi}
A.~Nakamura and K.~Odaka,
%``CONSTRUCTION OF LOCAL MESON OPERATOR FROM PATH ORDERED OPERATOR IN QCD in two-dimensions,''
Phys. Lett. B \textbf{105}, 392-396 (1981)
doi:10.1016/0370-2693(81)90786-3
%9 citations counted in INSPIRE as of 01 Oct 2024

%\cite{Rajeev:1994tr}
\bibitem{Rajeev:1994tr}
S.~G.~Rajeev,
%``Quantum hadrodynamics in two-dimensions,''
Int. J. Mod. Phys. A \textbf{9}, 5583-5624 (1994)
doi:10.1142/S0217751X94002284
[arXiv:hep-th/9401115 [hep-th]].
%46 citations counted in INSPIRE as of 25 Sep 2024

%\cite{Dhar:1994ib}
\bibitem{Dhar:1994ib}
A.~Dhar, G.~Mandal and S.~R.~Wadia,
%``String field theory of two-dimensional QCD: A Realization of W(infinity) algebra,''
Phys. Lett. B \textbf{329}, 15-26 (1994)
doi:10.1016/0370-2693(94)90511-8
[arXiv:hep-th/9403050 [hep-th]].
%43 citations counted in INSPIRE as of 19 Dec 2024

%\cite{Dhar:1994aw}
\bibitem{Dhar:1994aw}
A.~Dhar, P.~Lakdawala, G.~Mandal and S.~R.~Wadia,
%``String field theory of two-dimensional QCD on a cylinder: a realization of W(infinity) current algebra,''
Int. J. Mod. Phys. A \textbf{10}, 2189-2224 (1995)
doi:10.1142/S0217751X95001066
[arXiv:hep-th/9407026 [hep-th]].
%13 citations counted in INSPIRE as of 14 Oct 2024

%\cite{Cavicchi:1993jh}
\bibitem{Cavicchi:1993jh}
M.~Cavicchi,
%``A Bilocal field approach to the large N expansion of two-dimensional (gauge) theories,''
Int. J. Mod. Phys. A \textbf{10}, 167-198 (1995)
doi:10.1142/S0217751X95000097
[arXiv:hep-th/9401086 [hep-th]].
%19 citations counted in INSPIRE as of 14 Oct 2024

%\cite{Barbon:1994au}
\bibitem{Barbon:1994au}
J.~L.~F.~Barbon and K.~Demeterfi,
%``Effective Hamiltonians for 1/N expansion in two-dimensional QCD,''
Nucl. Phys. B \textbf{434}, 109-138 (1995)
doi:10.1016/0550-3213(94)00442-H
[arXiv:hep-th/9406046 [hep-th]].
%18 citations counted in INSPIRE as of 01 Oct 2024

%\cite{Itakura:1996bk}
\bibitem{Itakura:1996bk}
K.~Itakura,
%``Boson expansion methods in light front QCD in (1+1)-dimensions,''
Phys. Rev. D \textbf{54}, 2853-2862 (1996)
doi:10.1103/PhysRevD.54.2853
[arXiv:hep-th/9604032 [hep-th]].
%17 citations counted in INSPIRE as of 01 Oct 2024

%\cite{Collins:1981uw}
\bibitem{Collins:1981uw}
J.~C.~Collins and D.~E.~Soper,
%``Parton Distribution and Decay Functions,''
Nucl. Phys. B \textbf{194}, 445-492 (1982)
doi:10.1016/0550-3213(82)90021-9
%1264 citations counted in INSPIRE as of 18 Dec 2024

%\cite{Koepf:1994an}
\bibitem{Koepf:1994an}
W.~Koepf, E.~M.~Henley and M.~A.~Alberg,
%``Mesons and the structure of nucleons,''
doi:10.1142/9789812831408\_0024
[arXiv:nucl-th/9403014 [nucl-th]].
%1 citations counted in INSPIRE as of 09 Dec 2024

%\cite{Paiva:1996dd}
\bibitem{Paiva:1996dd}
S.~Paiva, M.~Nielsen, F.~S.~Navarra, F.~O.~Duraes and L.~L.~Barz,
%``Virtual meson cloud of the nucleon and intrinsic strangeness and charm,''
Mod. Phys. Lett. A \textbf{13}, 2715-2724 (1998)
doi:10.1142/S0217732398002886
[arXiv:hep-ph/9610310 [hep-ph]].
%68 citations counted in INSPIRE as of 28 Oct 2024

%\cite{Hobbs:2013bia}
\bibitem{Hobbs:2013bia}
T.~J.~Hobbs, J.~T.~Londergan and W.~Melnitchouk,
%``Phenomenology of nonperturbative charm in the nucleon,''
Phys. Rev. D \textbf{89}, no.7, 074008 (2014)
doi:10.1103/PhysRevD.89.074008
[arXiv:1311.1578 [hep-ph]].
%80 citations counted in INSPIRE as of 10 Dec 2024

%\cite{Melnitchouk:1997ig}
\bibitem{Melnitchouk:1997ig}
W.~Melnitchouk and A.~W.~Thomas,
%``HERA anomaly and hard charm in the nucleon,''
Phys. Lett. B \textbf{414}, 134-139 (1997)
doi:10.1016/S0370-2693(97)01150-7
[arXiv:hep-ph/9707387 [hep-ph]].
%75 citations counted in INSPIRE as of 19 Nov 2024

%\cite{Jia:2017uul}
\bibitem{Jia:2017uul}
Y.~Jia, S.~Liang, L.~Li and X.~Xiong,
%``Solving the Bars-Green equation for moving mesons in two-dimensional QCD,''
JHEP \textbf{11}, 151 (2017)
doi:10.1007/JHEP11(2017)151
[arXiv:1708.09379 [hep-ph]].
%36 citations counted in INSPIRE as of 01 Oct 2024

%\cite{Burkardt:2000uu}
\bibitem{Burkardt:2000uu}
M.~Burkardt,
%``Off forward parton distributions in (1+1)-dimensional QCD,''
Phys. Rev. D \textbf{62}, 094003 (2000)
doi:10.1103/PhysRevD.62.094003
[arXiv:hep-ph/0005209 [hep-ph]].
%31 citations counted in INSPIRE as of 01 Oct 2024

%\cite{Kalashnikova:2000dw}
\bibitem{Kalashnikova:2000dw}
Y.~S.~Kalashnikova and A.~V.~Nefediev,
%``Strong decays and Adler selfconsistency condition in two-dimensional QCD,''
Phys. Lett. B \textbf{487}, 371-378 (2000)
doi:10.1016/S0370-2693(00)00828-5
[arXiv:hep-ph/0006070 [hep-ph]].
%4 citations counted in INSPIRE as of 25 Sep 2024

\end{thebibliography}
\end{document}